\documentclass[sn-basic]{sn-jnl}% Math and Physical Sciences Reference Style
\usepackage{graphicx}%
\usepackage{multirow}%
\usepackage{amsmath,amssymb,amsfonts}%
\usepackage{amsthm}%
\usepackage{mathrsfs}%
\usepackage[title]{appendix}%
\usepackage{xcolor}%
\usepackage{textcomp}%
\usepackage{manyfoot}%
\usepackage{booktabs}%
\usepackage{algorithm}%
\usepackage{algorithmicx}%
\usepackage{algpseudocode}%
\usepackage{listings}%

\usepackage[utf8]{inputenc}
\usepackage{lipsum} % Package to generate dummy text throughout this template
\usepackage{blindtext}
\usepackage{graphicx} 
\usepackage{caption}
\usepackage[T1]{fontenc}
\usepackage{multicol} % Used for the two-column layout of the document
\usepackage{booktabs} % Horizontal rules in tables
\usepackage{float} % Required for tables and figures in the multi-column environment - they need to be placed in specific locations with the [H] (e.g. \begin{table}[H])
\usepackage{hyperref} % For hyperlinks in the PDF
\usepackage{paralist} % Used for the compactitem environment which makes bullet points with less space between them
\usepackage{natbib, url}
\usepackage{listings}
\usepackage{bm}
\usepackage{cancel}
\usepackage{comment}
\theoremstyle{thmstyleone}%
\theoremstyle{thmstyletwo}%

\theoremstyle{thmstylethree}%

\newcommand{\given}{\mid}
\newcommand{\T}{\mathrm{\scriptscriptstyle T}}

\begin{document}

\title[Bayesian Finite Population Survey Sampling]{Finite Population Survey Sampling: An Unapologetic Bayesian Perspective}


\author*[1]{\fnm{Sudipto} \sur{Banerjee}}\email{sudipto@ucla.edu}

%\author[2,3]{\fnm{Second} \sur{Author}}%\email{iiauthor@gmail.com}

%\author[1,2]{\fnm{Third} \sur{Author}}%\email{iiiauthor@gmail.com}

\affil*[1]{\orgdiv{UCLA Department of Biostatistics; UCLA Department of Statistics and Data Science}, \orgname{University of California Los Angeles}, \orgaddress{\street{650 Charles E. Young Drive South}, \city{Los Angeles}, \postcode{90095-1772}, \state{California}, \country{USA}}}

%\affil[2]{\orgdiv{Department}, \orgname{Organization}, \orgaddress{\street{Street}, \city{City}, \postcode{10587}, \state{State}, \country{Country}}}

%\affil[3]{\orgdiv{Department}, \orgname{Organization}, \orgaddress{\street{Street}, \city{City}, \postcode{610101}, \state{State}, \country{Country}}}


\abstract{This article attempts to offer some perspectives on Bayesian inference for finite population quantities when the units in the population are assumed to exhibit complex dependencies. Beginning with an overview of Bayesian hierarchical models, including some that yield design-based Horvitz-Thompson estimators, the article proceeds to introduce dependence in finite populations and sets out inferential frameworks for ignorable and nonignorable responses. Multivariate dependencies using graphical models and spatial processes are discussed and some salient features of two recent analyses for spatial finite populations are presented.}




\maketitle

\section{Introduction}\label{sec: intro}
\cite{basu1971} tells a delightful and impressionable tale of a circus statistician and an elephant to cogently illustrate, with characteristic humour, the potential pitfalls of the widely used design-based Horvitz-Thompson estimator \citep{narain1951, Horvitz:tp} in survey sampling. \cite{Little:2004jz}, in a splendidly written review paper, quotes this tale and offers substantial insights on inference for finite populations in an illuminating broader discussion of model-based and design-based approaches in survey sampling. These two paradigms have received significant attention in survey sampling and I do not intend to undertake a comprehensive review here. Finite population survey sampling has an extensive literature that is beyond the scope of a single article \citep[see classic texts such as][emphasizing design-based approaches]{Cochran:1977un, kish1965book}. Bayesian inference for finite population survey sampling has been discussed from diverse perspectives, for example, in \cite{Ericson:1969wi}, \cite{rao1972jasa}, \cite{arora1997empirical}, \cite{Ghosh:1997tp}, \cite{little2002book} \cite{Gelman:2007bh}, \cite{Little:2004jz} and \cite{Gelman:2014tc}, while an excellent appraisal of classical and Bayesian approaches is offered by \cite{rao2011statsci}. %\cite{Ericson:1969wi} is widely considered as one of the earliest accounts of Bayesian inference for finite populations and generated significant discussions concerning the role of the sampling design in Bayesian inference \citep{hartleyrao1968bioka, rao1972jasa}.  Numerous other manuscripts have considered inference for finite populations from different inferential perspectives, often adopting a stance somewhere between classical and Bayesian \citep[a small sample of notable articles include][]{hartleyrao1968bioka}
 \cite{DiZio2023} offer a wonderfully insightful review of Debabrata Basu's legacy in the development of survey sampling, which I will not undertake here. 

I will share some perspectives on Bayesian inference for finite population quantities with an emphasis on dependent finite populations. \cite{Little:2004jz} comments that advocating Bayesian inference for survey sampling is akin to ``swimming upstream'', given the aversion that many survey statisticians have to modelling assumptions, but notes the richness and flexibility available in models for improving design-based inference and for addressing survey weights. Advocates of design-based survey sampling, where inference is based on the randomisation distribution, have often criticised modelling assumptions such as the population units being ``independent and identically distributed'', which fails to account for the sampling design \citep[see, e.g.,][Section 9]{kish1995}. However, with computational resources available to statisticians today, building models for dependent finite population units yield rich and flexible classes of Bayesian models. A key point is that the usual model-based approaches (likelihood or Bayesian) often proceed from an assumption of ``ignorability'' that the analysis includes all variables that affect the probability of a unit being included in the sample. A different way to make this point is to say that the probability of obtaining a given sample does not depend upon the values of the population units. Under such assumptions, the likelihood, or the model of how the data was realised, drives the inference and the sampling design can be ignored \citep[see, e.g.,][for further discussions]{Rubin:1976we, Gelman:2007bh, Gelman:2014tc}.   

The Bayesian approach to statistical inference seeks to provide full uncertainty quantification of unknown quantities of interest conditional upon known (or seen) quantities. Denoting ${\cal U}$ to be a collection of unknowns and ${\cal K}$ a collection of ``knowns'', we wish to report the posterior distribution $[{\cal U} \given {\cal K}] \propto [{\cal U}, {\cal K}]$. Probabilistic inference arrives at this distribution by specifying $[{\cal K}\given {\cal U}]$ (the likelihood) and $[{\cal U}]$ (the prior). Treating Bayesian survey sampling from the perspective of missing data is appealing and conceptually simple. Here ${\cal K}$ represents the sampled data and ${\cal U}$ represents all unknowns including the population units not in the sample as well as parameters needed to specify a probability model for $[{\cal U}]$. The plethora of applications within this paradigm, especially in the context of complex surveys, is daunting due to sheer volume \cite[see, e.g.,][and references therein]{little1982models, little2002book, tang2003analysis, Gelman:2014tc, kalton2019isr}.  

Following conventional notations, let us assume that ${\cal N} = \{1,2,\ldots,N\}$ is a finite set of integers labelling the population units. Given a sample $\{i_1, i_2,\ldots,i_n\}$ of size $n$, which is a subset of ${\cal N}$, units in the population and sample are described by $\{(Y_t, Z_t) : t \in {\cal N}\}$, where $Y_t$ denotes the value of population unit labelled $t$ and $Z_t$ is a binary variable indicating if unit $t$ is in the sample or not. The binary vector $Z = (Z_1, Z_2,\ldots, Z_N)^{\T}$ completely encodes the sampled units in the population and, analogously, $Y = (Y_1, Y_2,\ldots, Y_N)^{\T}$ denotes the values of the all units in the finite population. Model-based inference (likelihood or Bayesian) assigns a probability law to the finite collection $(Y, Z)$, which defines the sampling design. It is sometimes convenient to imagine the exercise of sampling as a process over time where units enter the sample one by one. At time zero, then, a probability model for $Y$, $p(Y)$, can be interpreted as a prior distribution on the unknown population units. The sampling design determines the joint distribution over $Y$ and $Z$ by specifying $p(Z\given Y)$. Bayesian inference proceeds from $p(Y\given {\cal D}, Z) \propto p(Y) p(Z\given Y) p({\cal D}\given Y, Z)$, where ${\cal D}$ is the sampled data. 

In the domain of dependent finite population models, there is a substantial literature on small area estimation for regionally aggregated data \citep[see, e.g.,][]{Rao:vz, Clayton1987:eb, datta1991bayesian, Ghosh1994:sa, arora1997sinica, Ghosh1998:kn}, where we model dependencies across regions. If one considers ``being adjacent to'' or ``being a neighbour of'' as a relation over the set of regions, then one can build dependencies using a graph ${\cal G} = \{{\cal V}, {\cal E}\}$, where vertices in ${\cal V}$ correspond to the regions and the edges in ${\cal E}$ are incident upon pairs of vertices that represent adjacent regions. Graphical models recognise conditional dependence among adjacent nodes, and conditional independence otherwise, given all other nodes.

Much of this article revisits familiar models in Bayesian finite population inference, but attempts to offer additional insights from a more contemporary perspective of modelling complex multivariate dependencies. Section~\ref{sec: srswor_bhm} presents Bayesian inference for simple random sampling without replacement. Section~\ref{sec: bayes_ht} discusses Bayesian models that reproduce inference from Horvitz-Thompson estimators. Section~\ref{sec: bayesian_multistage} discusses Bayesian multistage sampling as devised in \cite{Scott:1969fp} and \cite{Malec:1985tr}, but with a modest generalisation to modelling unknown variances. I offer derivations in a systematic manner using conjugate Bayesian updating with minimal use of matrix computations. Section~\ref{sec: dep_pop} introduces models for dependent population units with an emphasis on spatial processes and multivariate dependencies. Ignorable and nonignorable responses are discussed in this context. Section~\ref{sec: model_assessment} briefly discusses model choice and assessment, while Section~\ref{sec: examples} presents some key results from analysis of finite population surveys in ignorable and nonignorable contexts. Section~\ref{sec: discussion} concludes the paper with a brief discussion.

\section{SRSWOR and Bayesian hierarchical models}\label{sec: srswor_bhm}

In simple random sampling without replacement (SRSWOR) from a population of $N$ units, $p(Z\given Y) = 1/\binom{N}{n}$ whenever $\sum_{i=1}^N Z_i = n$ and $0$ otherwise. Since this distribution does not depend upon $Y$, we can write $p(Y,Z) = p(Y)p(Z)$ and we seek the posterior distribution $p(Y\given {\cal D}, Z) \propto p(Y) p({\cal D}\given Y, Z)$. Inference on the population is based solely on the prior distribution for $Y$, which induces the distribution $p({\cal D}\given Y, Z)$. If the probability model assumes that the units are exchangeable, then we can reorder the labels as necessary and denote ${\cal D}$ by $y = (y_1, y_2,\ldots, y_n)^{\T}$ as the values of the $n$ sampled units and $Y_u = (Y_{n+1},\ldots, Y_N)^{\T}$ as the unknown values of units not sampled. As a familiar example, consider the simple hierarchical model for population units:
\begin{equation}
    \begin{split}
        Y_i \overset{iid}{\sim} N(\mu, \sigma^2),\; i=1,2,\ldots,N; \quad \mu \sim N(\theta, \sigma^2/n_0);\quad \sigma^2 \sim IG(a_0, b_0)\;,
    \end{split}
\end{equation}
where $\theta$, $n_0$, $a_0$ and $b_0$ are known scalars. The joint posterior factorises as
\begin{align*}
 p(\mu,\sigma^2 \given y) &\propto
 \begin{array}{ccc}
  \underbrace{IG(\sigma^2\given a_1, b_1)} & \times & \underbrace{N\left(\mu\,\left|\, \frac{n_0\theta + n\bar{y}}{n_0 + n}, \frac{\sigma^2}{n_0 + n}\right.\right)}\\
  p(\sigma^2\given y) & & p(\mu\given \sigma^2, y)
 \end{array}\; , 
\end{align*}
where $\bar{y} = \sum_{i=1}^n y_i/n$ is the sample mean, $ a_1 = a_0 + \frac{n}{2}$ and $b_1 = b_0 +  \frac{f(\theta,\bar{y})}{2}$, $f(\theta,\bar{y}) = (n-1)s^2 + \frac{n_0n}{n_0+n}(\bar{y} - \theta)^2$ and $s^2 = \sum_{i=1}^n (y_i - \bar{y})^2/(n-1)$. Conditional on $\sigma^2$, it is easy to derive the joint posterior distribution $p(Y_u, \mu \given \sigma^2, y) = N(\mu\given Mm, \sigma^2M) \times N(Y_u \given 1_{N-n}\mu, \sigma^2I_{N-n})$, where $m = n_0\theta + n\bar{y}$ and $M = 1/(n+n_0)$. The marginal posterior distribution $p(Y_u \given \sigma^2, y)$ is easily obtained (without requiring explicit integration) by substituting $\mu = Mm + \omega$, where $\omega \sim N(0, \sigma^2M)$, as $N(Y_u\given 1_{N-n}(n_0\theta + n\bar{y})/(n+n_0), \sigma^2(1_{N-n}1_{N-n}^{\T}/(n+n_0) + I_{N-n}/n))$. These calculations are straightforward in Bayesian statistics and are straightforward to adapt to finite population models. The posterior distribution for any linear function of the population units, say $p(a^{\T}Y\given \sigma^2, y)$, where $a=(a_1,\ldots,a_N)^{\T}$, is again Normal with mean
{%\footnotesize
\begin{equation}\label{eq: srswor_linear_posterior_mean}
%    \mathbb{E}[a^{\T}Y \given \sigma^2, y] = 
\sum_{i=1}^n a_iy_i + \sum_{i=n+1}^Na_i \mathbb{E}[Y_{u,i}\given \sigma^2, y] = \sum_{i=1}^n\left(a_i + \frac{\sum_{i=n+1}^N a_i}{n_0 + n}\right)y_i  + \frac{n_0\sum_{i=n+1}^N a_i}{n_0 + n}\theta\; .
\end{equation}
} 
The posterior covariance matrix of $Y_u$ is the sum of $\mathbb{E}[\mathbb{V}(Y_u \given \sigma^2, \mu, y)]$ and $\mathbb{V}(\mathbb{E}[Y_u \given \sigma^2, \mu, y])$, which yields $\sigma^2(1_{N-n}1_{N-n}^{\T}/(n+n_0) + I_{N-n})$. Therefore,
\begin{equation}\label{eq: srswor_linear_posterior_variance}
\mathbb{V}\left(a^{\T}Y_u \given \sigma^2, y\right) = \sigma^2 \left(\frac{\left(\sum_{i=n+1}^N a_i\right)^2}{n+n_0} + \sum_{i=n+1}^N a_i^2\right)\;.    
\end{equation}
For full inference when $\sigma^2$ is unknown, we first draw a sample of values of $\sigma^2 \sim IG(a_1, b_1)$ followed by one draw from a Normal distribution with mean and variance given by (\ref{eq: srswor_linear_posterior_mean}) and (\ref{eq: srswor_linear_posterior_variance}), respectively, for each drawn value of $\sigma^2$. 

For the population total, $a_i=1$ for each $i=1,2,\ldots,N$ so the mean and variance of $p(\sum_{i=1}^N Y_i\given \sigma^2, y)$ is given by $\sum_{i=1}^n (n_0 + N)y_i/(n_0 + n)$ and the variance is $\sigma^2\left((N-n)(n_0 + N)/(n_0+n)\right)$. Letting $f=n/N$ and $f_0=n_0/N$, the posterior distribution of the population mean, $p(\bar{Y}\given \sigma^2, y)$, can be expressed as
\begin{equation}\label{eq: posterior_population_mean}
    \left[\left.\frac{\bar{Y} - \left\{f\bar{y} + (1-f)\frac{n_0\theta + n\bar{y}}{n_0+n}\right\}}{\sigma\sqrt{\frac{(1-f)(1+f_0)}{n+n_0}}}\right|\sigma^2, y\right] \sim N(0,1)\;.
\end{equation}
If the prior distribution on $\mu$ becomes vague, i.e., $n_0\to 0$, then the posterior mean and variance in (\ref{eq: posterior_population_mean}) are $\bar{y}$ and $(1-f)\sigma^2/n$, respectively. Further, as $n_0\to 0$ and $a_0 \to -1/2$ (corresponds to a uniform prior over the real line for $(\mu, \log\sigma^2)$), the resulting posterior distribution $p(\sigma^2 \given y) \to IG((n-1)/2, (n-1)s^2/2)$ and the marginal posterior distribution for the population mean is given by $\left[\left.\frac{\bar{Y} - \bar{y}}{s\sqrt{(1-f)/n}}\,\right|\,y\right] \sim T_{n-1}$, where $T_{n-1}$ denotes the Student's $t$ distribution with $n-1$ degrees of freedom. This, numerically, reproduces the inference from the classical theory of design-based estimators while retaining the appealing interpretation of direct probability statements about the population mean and without recourse to the specialised mathematical developments of Central Limit Theorems for SRSWOR sampling from finite populations.   

\section{Bayesian and Horvitz-Thompson estimates}\label{sec: bayes_ht}

The posterior distribution of the population total, $N\bar{Y}$, has mean $\sum_{i=1}^n y_i/\pi_i$, where $\pi_i = n/N$ is the inclusion probability of the $i$-th unit in the sample. This is precisely the design-based Horvitz-Thompson estimator \citep{narain1951, Horvitz:tp} of the population total. This, then, begs the question whether there exists a Bayesian model that will yield a posterior mean equalling the Horvitz-Thompson estimator. The answer is in the affirmative and it is worth taking a look at a simple model \citep{Little:2004jz},
\begin{equation}\label{eq: bayesian_ht_little}
%   Y_i \overset{ind}{\sim} N(\pi_i\beta, \pi_i^2),\; i=1,2,\ldots,N;\quad p(\beta) \propto 1\;.  
Y_i = \pi_i \beta + \pi_i \epsilon_i;\quad \epsilon_i \overset{ind}{\sim} N(0,1),\; i=1,2,\ldots,N;\quad p(\beta) \propto 1\;,
\end{equation}
where each $\pi_i$ denotes the probability of population unit $i$ being included in the sample. For now, assume that $\pi_i$'s are known. Again, without loss of generality and appropriate relabelling of the population units, we assume that the labels $i=1,2,\ldots,n$ represent units that have been sampled. The posterior distribution $p(\beta \given y)$ is Normal with mean
\[
\mathbb{E}[\beta\given y] = (\pi^{\T}\mbox{diag}(1/\pi_i^2)\pi)^{-1}\pi^{\T}\mbox{diag}(1/\pi_i^2)y = (1/n)\sum_{i=1}^n (y_i/\pi_i)\;, 
\]
where $\pi = (\pi_1,\ldots,\pi_n)^{\T}$ and $\mbox{diag}(1/\pi_i^2)$ is the $n\times n$ diagonal matrix with elements $1/\pi_i^2$. Therefore, the posterior mean of the population total is
\begin{equation}\label{eq: bayesian_ht_little_derivation}
\begin{split}
    \mathbb{E}\left[{\left.\sum_{i=1}^N Y_i \,\right|\,} y \right] &= \sum_{i=1}^n y_i + \sum_{i=n+1}^N \mathbb{E}[Y_i \given y] = \sum_{i=1}^n y_i + \sum_{i=n+1}^N \pi_i\mathbb{E}[\beta \given y] \\
    &= \sum_{i=1}^n y_i + \left(\sum_{i=n+1}^N \pi_i\right)\mathbb{E}[\beta \given y] = \sum_{i=1}^n y_i + \left(n - \sum_{i=1}^n \pi_i\right)\mathbb{E}[\beta \given y] \\
    &= n\mathbb{E}[\beta \given y] + \sum_{i=1}^n \left(y_i - \pi_i\mathbb{E}[\beta\given y]\right) = \sum_{i=1}^n(y_i/\pi_i) + \sum_{i=1}^n \left(y_i - \hat{y}_i\right)\;,
\end{split}    
\end{equation}
where we have used the fact that $\sum_{i=1}^N \pi_i = \sum_{i=1}^N \mathbb{E}[Z_i] = \mathbb{E}\left[\sum_{i=1}^N Z_i\right] = n$ in the fourth equality and $\hat{y}_i = \mathbb{E}[\pi_i\beta\given y] = (\pi_i/n)\sum_{j=1}^n (y_j/\pi_j)$ in the last equality. The first term in the last expression is the Horvitz-Thompson estimator, while the second term is a sum of the differences between the observed and model-predicted values of the sampled units. Letting $n$ approach the population size, i.e., if the entire population is sampled, the second term vanishes and the posterior mean is the Horvitz-Thompson estimator. This is a model where the Bayesian estimate of the population total approximates the Horvitz-Thompson estimator.  

To derive a model where the posterior expectation of the population total is \emph{exactly} the Horvitz-Thompson estimator, we modify (\ref{eq: bayesian_ht_little}) as follows,
\begin{equation}\label{eq: bayesian_ht_sinha_ghosh}
%   Y_i \overset{ind}{\sim} N(\pi_i\beta, \pi_i^2),\; i=1,2,\ldots,N;\quad p(\beta) \propto 1\;.  
Y_i = \pi_i \beta + \pi_i \epsilon_i;\quad \epsilon_i \overset{ind}{\sim} N(0,\sigma_i^2),\; i=1,2,\ldots,N;\quad p(\beta) \propto 1\;,
\end{equation}
where the $\sigma_i^2$'s are assumed to be fixed constants. The posterior mean of $\beta$ is
\[
\mathbb{E}[\beta \given y] = (\pi^{\T}\mbox{diag}(1/(\sigma_i\pi_i)^2)\pi)^{-1}\pi^{\T}\mbox{diag}(1/(\sigma_i\pi_i)^2)y 
= \sum_{i=1}^n w_i(y_i/\pi_i)\;,
\]
where $w_i = (1/\sigma_i^2)/\left(\sum_{i=1}^n 1/\sigma_i^2\right)$. Proceeding analogous to (\ref{eq: bayesian_ht_little_derivation}), we again have 
\begin{equation}
    \begin{split}
    \mathbb{E}\left[\left.\sum_{i=1}^N Y_i \,\right|\, y\right] &= \sum_{i=1}^n y_i + \left(\sum_{i=n+1}^N \pi_i\right)\mathbb{E}[\beta \given y] = \sum_{i=1}^n y_i + \left(\sum_{i=n+1}^N \pi_i\right)\left(\sum_{i=1}^n w_i(y_i/\pi_i)\right) \\
    &= \sum_{i=1}^n\left(1 + \left(n - \sum_{i=1}^n\pi_i\right)w_i/\pi_i\right)y_i = \sum_{i=1}^n \tilde{w}_i y_i\;, 
    \end{split}
\end{equation}
where $\tilde{w}_i = 1 + \left(n-\sum_{i=1}^n\pi_i\right)w_i/\pi_i$. We will obtain the Horvitz-Thompson estimator in the last expression if $\tilde{w}_i = 1/\pi_i$. Letting $\sigma_i^2 = 1/(1-\pi_i)$ yields $w_i = (1-\pi_i)/\left(n-\sum_{i=1}^n\pi_i\right)$ so that $\tilde{w}_i = 1 + (1-\pi_i)/\pi_i = 1/\pi_i$. Therefore, setting $\sigma_i^2 = 1/(1-\pi_i)$ in (\ref{eq: bayesian_ht_sinha_ghosh}) yields a Bayesian model where the posterior expectation of the population total exactly agrees with the Horvitz-Thompson estimator \citep{Ghosh:1990ht}. See \cite{Ghosh:2012md} for generalisations of this result to exponential families and other families of priors. 

\section{Bayesian multistage sampling}\label{sec: bayesian_multistage}
Multistage sampling designs draw samples from the population in multiple stages. The population is assumed to be composed of primary (level 1) units from which a random sample is drawn using SRSWOR. Next, for each of these sampled primary units, a random sample of secondary (level 2) units is drawn and, if needed, this process is continued to further sample from the secondary units. Consider 2-stage sampling. Let $i=1,2,\ldots,N$ be the labels for the primary population units and let $j=1,2,\ldots,M_i$ be the number of secondary units within the $i$-th primary unit. Consider a sampling design that draws $n$ primary units using SRSWOR and, then, for each of these primary units draws a sample of $m_i$ secondary units using SRSWOR. If $Y_{(i,j)}$ denotes the population unit referenced by secondary unit $j$ within primary unit $i$, then the Horvitz-Thompson estimator for the population total is $\sum_{i,j=1}^{n,m_i} y_{(i,j)}/\pi_{i,j}$, where $\pi_{i,j} = (n/N)(m_i/M_i) = P(Z_{(i,j)} = 1)$, where $Z_{(i,j)} = 1$ if the secondary unit $j$ within primary unit $i$ is sampled and $Z_{(i,j)}=0$ otherwise. 

\subsection{Bayesian hierarchical model for 2-stage sampling}\label{subsec: bhm_two_stage}

A Bayesian model for 2-stage sampling reproducing the Horvitz-Thompson estimator as its posterior mean is easily derived from (\ref{eq: bayesian_ht_sinha_ghosh}) by writing $Y_{(i,j)} = \pi_{(i,j)}\beta + \pi_{(i,j)}\epsilon_{(i,j)}$, where $\epsilon_{(i,j)} \overset{ind}{\sim} N(0, 1/(1-\pi_{(i,j)}))$. However, richer classes of models arise from Bayesian hierarchical models for 2-stage sampling \citep{Scott:1969fp}, while \cite{Malec:1985tr} discuss Bayesian inference for more general multistage sampling. Here, we will take a closer look at the 2-stage model but in a more contemporary context of ``BIG DATA'' settings, where the CPU storage and memory allocations will not allow us to deal with data of size exceeding $O(\max_{1\leq i \leq N} m_i)$. Bayesian analysis will, therefore, have to proceed sequentially as samples stream in using a ``divide and conquer'' mechanism.

Let us turn to a 2-stage sampling design in the context of a finite population of massive size, where resources allow us to sample only $n$ out of $N$ population units and data of size no more than $\max_{1\leq i \leq n} m_i$ units for analysis. We consider the model
\begin{equation}\label{eq: bayesian_two_stage}
    \begin{split}
        Y_{ij} &\overset{ind}{\sim} N(\beta_i, \delta^2/\gamma_i^2),\; j=1,2,\ldots,M_i; \\
        \beta_i &\overset{ind}{\sim} N(\nu, \delta^2),\; i=1,2,\ldots,N;\\
        \nu &\sim N(c_0, \delta^2/n_0);\quad \delta^2 \sim IG(a_0, b_0)\;,
    \end{split}
\end{equation}
where $\gamma_i^2$, $n_0$, $c_0$, $a_0$ and $b_0$ are fixed scalars. Suppose, with legitimate relabelling of the population units that is allowed due to exchangeable random variables in (\ref{eq: bayesian_two_stage}), we sample $i=1,2,\ldots,n$ primary units from the population and $j=1,2,\ldots,m_i$ secondary units from the $i$-th primary unit. Let $y_i = (y_{i1},\ldots,y_{im_i})^{\T}$ be the $m_i\times 1$ vector of sampled observations from the $i$-th primary unit in the population. The first equation in (\ref{eq: bayesian_two_stage}) yields $y_i = 1_{m_i}\beta_i + \epsilon_i$, where $1_{m_i}$ is the $m_i\times 1$ vector of ones, $\beta_i$ is the scalar parameter corresponding to the mean of the $i$-th primary unit, and $\epsilon_i \overset{ind}{\sim} N\left(0, (\delta^2/\gamma_i^2)I_{m_i}\right)$ with $I_{m_i}$ denoting the $m_i\times m_i$ identity matrix. Substituting the distribution of $\beta_i$ in the second equation of (\ref{eq: bayesian_two_stage}) results in integrating out $\beta_i$ from the model and we obtain $p(y_i\given \nu, \delta^2)$ as the model $y_i = 1_{m_i}\nu + \omega_i$, where $\omega_i \overset{ind}{\sim} N\left(0, \delta^2V_i\right)$, where $V_i = (1/\gamma_i^2)I_{m_i} + 1_{m_i}1_{m_i}^{\T}$. To optimise storage and computations we follow sequential Bayesian updating where the posterior distribution is updated using data from one primary unit only at each iteration. The resulting posterior distribution then acts as the prior distribution to be updated using data from the next primary unit. 

\subsection{Inference for $\{\nu, \delta^2\}$}\label{subsec: bhm_nu_delta}

To illustrate in the context of (\ref{eq: bayesian_two_stage}), we first derive the posterior distribution $p(\nu, \delta^2 \given y_1) \propto IG(\delta^2\given a_0, b_0)\times N(\nu\given c_0, \delta^2/n_0) \times N\left(y_1\given 1_{m_i}\nu, \delta^2V_1\right)$, which, after standard Bayesian Normal-Normal calculations, yields
\begin{equation*}
\begin{split}
 p(\nu,\delta^2 \given y_1) &=
 \begin{array}{ccc}
  \underbrace{IG(\delta^2\given a_{1}, b_{1})} & \times & \underbrace{N\left(\nu\,\left|\, C_1c_1, \delta^2 C_1\right.\right)}\\
  p(\delta^2\given y_1) & & p(\nu\given \delta^2, y_1)
 \end{array}\; ,
 \end{split}\end{equation*}
where $c_1 = n_0c_0 + 1_{m_1}^{\T}V_1^{-1}y_1$, $C_1^{-1} = n_0 + 1_{m_1}^{\T}V_1^{-1}1_{m_1}$, $a_1 = a_0 + m_1/2$ and $b_1 = b_0 + q_1/2$ with $q_1 = n_0c_0^2 + y_1^{\T}V_1^{-1}y_1 - c_1^{2}C_1$. Simplifications arise from applying the Sherman-Morrison matrix identity to obtain
\begin{equation}\label{eq: bayesian_two_stage_swm_app}
V_1^{-1} = \gamma_1^2 I_{m_1} - \frac{\gamma_1^2 1_{m_1}1_{m_1}^{\T}\gamma_1^2}{1 + m_1\gamma_1^2} = \gamma_1^2\left(I_{m_1} - \frac{\gamma_1^2}{1+m_1\gamma_1^2}1_{m_1}1_{m_1}^{\T}\right). %\\
%= \frac{\lambda_1}{m_1(1-\lambda_1)}\left(I_{m_1} - \frac{\lambda_1}{m_1}1_{m_1}1_{m_1}^{\T}\right) = \frac{\lambda_1}{m_1(1-\lambda_1)}I_{m_1} - \frac{\lambda_1^2}{m_1^2(1-\lambda_1)}1_{m_1}1_{m_1}^{\T},
\end{equation}
%where 
Writing %
$\lambda_1 = m_1\gamma_1^2/(1+m_1\gamma_1^2)$%. W
, 
we obtain the following simplified expressions: $c_1 = n_0c_0 + \lambda_1\bar{y}_1$ and $C_1^{-1} = n_0 + \lambda_1$, which yields $\mathbb{E}[\nu \given y_1] = n_0/(n_0+\lambda_1)c_0 + \lambda_1/(n_0 + \lambda_1)\bar{y_1}$. Treating this as the base case, we apply induction to conclude for each $t=1,2,\ldots,n$ 
\begin{equation}\label{eq: bayesian_two_stage_posterior}
    p(\nu, \delta^2 \given y_1,\ldots,y_t) = \begin{array}{ccc}
  \underbrace{IG(\delta^2\given a_{t}, b_{t})} & \times & \underbrace{N\left(\nu\,\left|\, C_tc_t, \delta^2 C_t\right.\right)}\\
  p(\delta^2\given y_1,\ldots,y_t) & & p(\nu\given \delta^2, y_1,\ldots,y_t)
 \end{array}\; ,
\end{equation}
where $c_t = c_{t-1} + 1_{m_t}^{\T}V_t^{-1}y_t$, $C_t^{-1} = C_{t-1}^{-1} + 1_{m_t}^{\T}V_t^{-1}1_{m_t}$, $a_t = a_{t-1} + m_t/2$ and $b_t = b_{t-1} + q_t/2$ with $q_t = c_{t-1}^2C_{t-1} + y_t^{\T}V_t^{-1}y_t - c_t^{2}C_t$. Applying (\ref{eq: bayesian_two_stage_swm_app}) to $V_t^{-1}$ again yields $c_t = c_{t-1} + \lambda_t\bar{y}_t$ and $C_{t}^{-1} = C_{t-1}^{-1} + \lambda_t$, where $\lambda_t = m_t\gamma_t^2/(1+m_t\gamma_t^2)$ for each $t=1,\ldots,n$, we obtain the expression
\begin{equation}\label{eq: bayesian_two_stage_nu_mean}
    \mathbb{E}[\nu\given \delta^2, y_1,\ldots,y_t] = \frac{n_0}{n_0 + \sum_{i=1}^t\lambda_i}c_0 + \sum_{i=1}^t\frac{\lambda_i}{n_0 + \sum_{i=1}^t\lambda_i}\bar{y}_i\;.
\end{equation}
Since $\mathbb{E}[\nu\given \delta^2, y_1,\ldots,y_t]$ does not depend on $\delta$, (\ref{eq: bayesian_two_stage_nu_mean}) is also the value of the marginal posterior mean $\mathbb{E}[\nu\given y_1,\ldots,y_t]$. Furthermore, $\mathbb{V}\left(\mathbb{E}[\nu\given \delta^2, y_1,\ldots,y_t]\right) = 0$ with respect to the density $p(\delta^2\given y_1,\ldots,y_t)$ and, therefore, the marginal posterior variance of $\nu$ is
\begin{equation}\label{eq: bayesian_two_stage_nu_variance}
    \mathbb{V}(\nu\given y_1,\ldots,y_t) %= \mathbb{E}\left[\mathbb{V}\left(\nu \given \delta^2, y_1,\ldots,y_t\right)\right] 
    = \frac{b_t}{a_{t}-1}C_t %\\ 
    = \frac{b_0 + \sum_{i=1}^t q_i/2}{\left(a_0+\sum_{i=1}^tm_i/2 - 1\right)\left(n_0 + \sum_{i=1}^t\lambda_i\right)}\;.
\end{equation}

\subsection{Inference for $\{\beta_1,\ldots,\beta_N\}$}\label{subsec: bhm_beta}

Turning to inference for $\beta = (\beta_1, \beta_2, \ldots, \beta_N)^{\T}$, note that 
\begin{equation*}\label{eq: bayesian_two_stage_beta1_posterior}
    \begin{split}
    p(\beta_t \given \nu, \delta^2, y_t) &\propto N(\beta_t \given \nu, \delta^2) \times N(y_t \given 1_{m_t}\beta_t, \delta^2/\gamma_t^2) \\
    &= N(\beta_t \given C_{\beta_t}c_{\beta_t}, \delta^2C_{\beta_t}) = N(\beta_t \given (1-\lambda_t)\nu + \lambda_t\bar{y}_t, \delta^2(1-\lambda_t))\;
    \end{split}
\end{equation*}
where the last equality follows from $c_{\beta_t} = \nu + \gamma_1^2m_t\bar{y}_t = \nu + \left(\lambda_t/(1-\lambda_t)\right)\bar{y}_t$ and $C_{\beta_1}^{-1} = 1 + m_t\gamma_t^2 = 1/(1-\lambda_t)$. The posterior distribution $p(\beta_1,\ldots,\beta_t\given \nu, \delta^2, y_1,\ldots,y_t)$ for each $t=1,\ldots,n$ is proportional to {\small
\begin{equation}\label{eq: bayesian_two_stage_posterior_beta_cond}
\begin{split}
&\begin{array}{ccccc}    
\underbrace{\prod_{i=1}^{t-1}N(\beta_i\given 1-\lambda_i)\nu + \lambda_i\bar{y}_i, \delta^2(1-\lambda_i))} & \times & \underbrace{N(\beta_t \given \nu, \delta^2)} & \times & \underbrace{N(y_t \given \beta_t, \delta^2\gamma_t^2)} \\
p(\beta_1,\ldots,\beta_{t-1} \given \nu, \delta^2, y_1,\ldots,y_{t-1}) & & p(\beta_t\given \nu, \delta^2) & & p(y_t\given \beta_t, \delta^2)
\end{array}\\
&\propto \begin{array}{ccc}
\underbrace{\prod_{i=1}^{t-1}N(\beta_i\given 1-\lambda_i)\nu + \lambda_i\bar{y}_i, \delta^2(1-\lambda_i))} & \times & \underbrace{N(\beta_t \given (1-\lambda_t)\nu + \lambda_t\bar{y}_t, \delta^2(1-\lambda_t))} \\
p(\beta_1,\ldots,\beta_{t-1} \given \nu, \delta^2, y_1,\ldots,y_{t-1}) & & p(\beta_t \given \nu, \delta^2, y_t)
\end{array}\;.
\end{split}
\end{equation} } 
Therefore, $p(\beta_1,\ldots,\beta_t\given \nu, \delta^2, y_1,\ldots,y_t) = \prod_{i=1}^t N(\beta_i\given (1-\lambda_i)\nu + \lambda_i\bar{y}_i, \delta^2(1-\lambda_i))$ for each $t=1,2,\ldots,n$. For $t = n+1,\ldots,N$, the primary units have not been sampled and, hence, there is no data available to update the priors. Hence, we can write
\begin{equation}\label{eq: bayesian_two_stage_beta_posterior}
    p(\beta \given \nu, \delta^2, y_1,\ldots, y_n) = \prod_{i=1}^N N\left(\beta_i \given (1-\lambda_i)\nu + \lambda_i\bar{y}_i, \delta^2(1-\lambda_i)\right)\;,
\end{equation}
where we define $m_i = 0$ and, hence, $\lambda_i = 0$, for $i=n+1,\ldots,N$. The marginal posterior distribution $p(\beta \given \delta^2, y_1,\ldots,y_n)$ is easily obtained by writing (\ref{eq: bayesian_two_stage_beta_posterior}) as $\beta = (1_{N}-\lambda)\nu + \Lambda\bar{y} + \omega$, where $1_{N}-\lambda = (1-\lambda_1,\ldots,1-\lambda_N)^{\T}$ is $N\times 1$, $\Lambda = \mbox{diag}(\lambda_1,\ldots,\lambda_N)$ is the $N\times N$ diagonal matrix with elements $1-\lambda_i$, $\bar{y} = (\bar{y}_1,\ldots,\bar{y}_n,0,\ldots,0)^{\T}$ is $N\times 1$ with the last $N-n$ terms being arbitrarily set to $0$ and $\omega \sim N(0, \delta^2(I_N - \Lambda))$. Therefore,
\begin{equation}\label{eq: bayesian_two_stage_beta_posterior_marginal}
    \beta \given \delta^2, y_1,\ldots, y_n \sim N\left((1_N-\lambda)\bar{y}_{.} + \Lambda\bar{y}, \delta^2(I_N-\Lambda) + \eta^2(1_N-\lambda)(1_N-\lambda)^{\T})\right)\;,
\end{equation}
where $\bar{y}_{.} = \mathbb{E}[\nu \given \delta^2, y_1,\ldots, y_n]$ given in (\ref{eq: bayesian_two_stage_nu_mean}) and $\eta^2 = \delta^2/(n_0 + \sum_{i=1}^N\lambda_i)$.

We now have the necessary distributions to sample from the posterior distribution of the model parameters $\{\delta^2, \nu, \beta\}$. We draw a sample from $p(\nu,\delta^2\given y_1,\ldots,y_n)$ as described earlier and draw a value of $\beta$ from (\ref{eq: bayesian_two_stage_beta_posterior}) for each drawn value of $\{\delta^2,\nu\}$. The resulting samples of $\beta$ are draws from $p(\beta \given y_1,\ldots,y_n)$. Note that this is computationally more efficient than drawing samples of $\beta$ from (\ref{eq: bayesian_two_stage_beta_posterior_marginal}) for each drawn $\delta^2$ since drawing from (\ref{eq: bayesian_two_stage_beta_posterior_marginal}) will require expensive matrix computations involving the $N\times N$ covariance matrix, where $N$ may exceed our data processing capabilities.

\subsection{Inference for all unobserved population units}\label{subsec: bhm_Y}

We now turn to inference for the unobserved population units. Let $Y_{i} = (Y_{i,m_i+1},\ldots, Y_{i,M_i})^{\T}$ for each $i=1,2,\ldots,N$ and let $Y_u = (Y_1^{\T},\ldots, Y_N^{\T})^{\T}$ be the vector of all unobserved population units, where we are using the definition $m_{i} = 0$ for $i=n+1,\ldots,N$. The random variable $Y_u$ is independent of the observed data given $\{\beta,\delta^2\}$ and its full conditional distribution is 
\begin{equation}\label{eq: posterior_predictive_unsampled_units}
    p(Y_u \given \beta, \delta^2) = \prod_{i=1}^N N\left(Y_i \given 1_{M_i-m_i}\beta_i, (\delta^2/\gamma^2_i)I_{M_i-m_i}\right)\;.
\end{equation}
Drawing samples from the posterior distribution $p(Y_{ij}\given y_1,\ldots,y_n) = \int p(Y_{ij}\given \beta_i, \delta^2)\times p(\beta_i, \delta^2 \given y_1,\ldots,y_n)$ for any unobserved population unit $(i,j)$ now proceeds by drawing one value of $Y_{ij} \sim N(\beta_i, \delta^2/\gamma_i^2)$ for each value of the posterior samples of $\beta_i$ and $\delta^2$. Again, no matrices are involved in this sampling step.   

Posterior inference for finite population quantities such as $\Omega = \sum_{i=1}^N\sum_{j=1}^{M_i} a_{ij}Y_{ij}$ is straightforward. We split up the sum as $\sum_{i=1}^n\sum_{j=1}^{m_i} a_{ij}y_{ij} + \sum_{i=1}^N\sum_{j=m_i+1}^{M_i} a_{ij}Y_{ij}$ and obtain posterior samples of this quantity by simply substituting the sampled values of $Y_{ij}$ in the second sum. Explicit algebraic expressions for the posterior mean and variance are not required for computations. Nevertheless, we have all the ingredients to obtain these moments. Let $a_i = \sum_{j=m_i+1}^{M_i} a_{ij}$ and recall that $\lambda_i=0$ for $i=n+1,\ldots,n$. Since $\mathbb{E}\left[Y_{ij}\given y_1,\ldots,y_n\right] = \mathbb{E}\left[\beta_{i}\given y_1,\ldots,y_n\right] = (1-\lambda_i)\bar{y}_{.} + \lambda_i\bar{y}_i$, the finite population estimate of $\Omega$ in the limit as $n_0 \to 0$ in (\ref{eq: bayesian_two_stage_nu_mean}) \citep[given in][]{Scott:1969fp} is obtained as
\begin{equation}\label{eq: scott_smith_estimate}
    \begin{split}
 &       \mathbb{E}\left[\Omega\given y_1,\ldots, y_n\right] = \sum_{i=1}^n\sum_{j=1}^{m_i} a_{ij}y_{ij} + \sum_{i=1}^N\sum_{j=m_i+1}^{M_i} a_{ij}\mathbb{E}\left[Y_{ij}\given y_1,\ldots,y_n\right] \\
        &\qquad\quad = \sum_{i=1}^n\sum_{j=1}^{m_i} a_{ij}y_{ij} + \sum_{i=1}^N a_i\{(1-\lambda_i)\bar{y}_{.} + \lambda_i\bar{y}_i\} \\
        &\qquad\quad = \sum_{i=1}^n\sum_{j=1}^{m_i} a_{ij}y_{ij} + \sum_{i=1}^N a_i\lambda_i\bar{y}_i + \left(\sum_{i=1}^N a_i(1-\lambda_i)\right)\bar{y}_{.} \\ 
        &\qquad\quad = \sum_{i=1}^n\sum_{j=1}^{m_i} a_{ij}y_{ij} + \sum_{i=1}^n \frac{a_i\lambda_i}{m_i}\sum_{j=1}^{m_i}y_{ij} + \frac{\sum_{i=1}^N a_i(1-\lambda_i)}{\sum_{i=1}^N\lambda_i}\sum_{i=1}^n\frac{\lambda_i}{m_i}\sum_{j=1}^{m_i}y_{ij} \\
        &\qquad\quad =  \sum_{i=1}^n\sum_{j=1}^{m_i}\left\{a_{ij} + a_i\frac{\lambda_i}{m_i} + \frac{\sum_{i=1}^N a_i(1-\lambda_i)}{\sum_{i=1}^N\lambda_i}\frac{\lambda_i}{m_i}\right\}y_{ij} \\
         &\qquad\quad =  \sum_{i=1}^n\sum_{j=1}^{m_i}\left\{a_{ij} + (a_i+k)\frac{\lambda_i}{m_i}\right\}y_{ij},\; \mbox{ where }\; k= \frac{\sum_{i=1}^N a_i(1-\lambda_i)}{\sum_{i=1}^N\lambda_i}\;.
    \end{split}
\end{equation}
The posterior estimate of the population total is obtained from (\ref{eq: scott_smith_estimate}) by setting $a_{ij} =1$, which implies $a_i = M_i-m_i$. The relationship between the Bayesian estimate and several design-based estimators is explored in \cite{Scott:1969fp}. The Horvitz-Thompson estimator does not emerge naturally from the posterior mean. Nevertheless, some simplifications are possible. If we further assume that $m_i$ and $\gamma_i^2$ vary so that $\lambda_i = \lambda$ is constant for each primary unit, then (\ref{eq: scott_smith_estimate}) yields
\begin{equation*}
    \sum_{i=1}^n\sum_{j=1}^{m_i} \left\{1-\lambda + \lambda\frac{M_i}{m_i} + (1-\lambda)\frac{\bar{a}}{m_i} \right\}y_{ij} = \lambda\sum_{i=1}^n M_i\bar{y}_i + (1-\lambda)\sum_{i=1}^n\left\{1 + \frac{\bar{a}}{m_i}\right\}m_i\bar{y_i} \;,
\end{equation*}
where $\bar{a} = \sum_{i=1}^N a_i / N$. In the limit as $\lambda \to 1$, the above expression equals $\sum_{i=1}^n M_i\bar{y}_i$. Hence, a simple modification of (\ref{eq: bayesian_two_stage}) with $Y_{ij} \overset{ind}{\sim} N(x_{i}\beta_i, \delta^2/\gamma_i^2)$ and each $x_i = n/N$ for $i=1,2,\ldots,N$ will yield a posterior mean $(N/n)\sum_{i=1}^n M_i\bar{y}_i$ in the limit as $\lambda \to 1$, which is precisely the Horvitz-Thompson estimator.  

\subsection{Conjugate Bayesian regression in ``BIG DATA'' settings}\label{subsec: bayesian_regression_big_data}

It is worth remarking, more generally, that conjugate Bayesian linear regression analysis renders itself as an extremely viable option for massive data sets. If the size of the data exceeds available memory capabilities and cannot be accessed in its entirety due to memory allocations, we split up the data into $n$ mutually exclusive blocks, each of size $m_i$, so that $\sum_{i=1}^n m_i$ is the total number of observations. This partitioning of the data is performed in a manner so that our memory resources are capable of analysing the individual data blocks. For example, consider the model   
\begin{equation}\label{eq: conjugate_bayesian_linear_regression_blocks}
    y_i \overset{ind}{\sim} N(X_i\beta, \sigma^2V_i),\; i=1,\ldots, n;\quad \beta \given \sigma^2 \sim N(C_0c_0, \sigma^2C_0);\quad \sigma^2 \sim IG(a_0, b_0)\;,
\end{equation}
where each $y_i$ is $m_i\times 1$, $X_i$ is an $m_i\times p$ matrix consisting of the values of known explanatory variables, $V_i$ is an $m_i\times m_i$ positive definite covariance matrix, $\beta$ is the $p\times 1$ vector of slopes that is assigned a $p$-variate multivariate Gaussian prior specified using a $p\times p$ matrix $C_0$ and $p\times 1$ vector $c_0$ and $\sigma^2$ is the residual variance. Sequential Bayesian updating uses the simple observation that $p(\beta, \sigma^2 \given y_1,\ldots, y_{t}) \propto p(\beta, \sigma^2 \given y_1,\ldots,y_{t-1}) \times p(y_t \given \beta, \sigma^2)$. This yields the joint posterior for each $t=1,2,\ldots,n$ as
\begin{equation}\label{eq: bayes_posterior_initial}
\begin{split}
 p(\beta,\sigma^2 \given y_1,\ldots, y_t) &=
 \begin{array}{ccc}
  \underbrace{IG(\sigma^2\given a_{t}, b_{t})} & \times & \underbrace{N\left(\beta\,\left|\, C_tc_t, \sigma^2 C_t\right.\right)}\\
  p(\sigma^2\given y_1,\ldots,y_t) & & p(\beta\given \sigma^2, y_1,\ldots,y_t)
 \end{array}\; ,
 \end{split}
\end{equation}
where $a_t = a_{t-1} + m_{t}/2$, $b_t = b_{t-1} + q_t/2$, $q_t = y_t^{\T}V_t^{-1}y_{t} + c_{t-1}^{\T}C_{t-1}c_{t-1} - c_t^{\T}C_tc_t$, $c_t = c_{t-1} + X_t^{\T}V_t^{-1}y_t$ and $C_t^{-1} = C_{t-1}^{-1} + X_t^{\T}V_{t}^{-1}X_t$. Since each step now involves only storage and computation in the order of at most $O(m^3)$ flops, where $m=\max_{1\leq t \leq n} m_t$, we obtain our desired posterior distribution at $t=n$ by applying (\ref{eq: conjugate_bayesian_linear_regression_blocks}) to one block of data in each iteration and overwriting these objects as we move to the next iteration. Posterior inference for an unobserved population unit $Y_{ij}$ that belongs to block $i$ will only access information from that block. For each sampled values $\beta$ and $\sigma^2$, we will draw one value $Y_{ij} \sim N(X_i\beta, \sigma^2 V_i)$, which results in a draw from $p(Y_{ij} \given y_1,\ldots,n)$. 

\subsection{Extensions}\label{subsec: extensions_multistage}
Extending the above models to three or higher stages is straightforward, although the algebra becomes more tedious. In practice, we can conveniently build hierarchical multistage models in programming environments such as \texttt{BUGS} \citep{bugs2009}, \texttt{rstan} \citep{rstan} and \texttt{nimble} \citep{nimble2017jcgs, nimbe2023package}. As we discussed in the preceding section, the conditional independence across primary units accrue significant computational benefits when dealing with large, or even massive, data sets since we can exploit the sequential nature of Bayesian updating in \eqref{eq: bayes_posterior_initial}. Section~\ref{sec: dep_pop} turns to multivariate models to infer on populations with spatially dependent units.

\section{Dependent population models}\label{sec: dep_pop}
A key advantage of the model-based approach to survey sampling is its ability to incorporate information, either known or posited, about the population units. Subsequent inference about the finite population quantities will follow from the probability distribution of unobserved quantities while accounting for the dependence structure in the population. Here, it will be instructive to recall the modelling framework promulgated by \cite{Rubin:1976we}; also see \cite{Gelman:2014tc}. As in Section~\ref{sec: intro}, let $Y = (y_1,\ldots,y_N)^{\T}$ be the units in the population and let $Z = (Z_1,\ldots,Z_N)^{\T}$ be the sampling inclusion indicators. A probability model for the entire finite population is given by $p(Y, Z \given \theta, \phi) = p(Y \given \theta) \times p(Z\given Y, \phi)$, where $\theta$ and $\phi$ are unknown parameters specifying the distribution of $Y$ and the conditional distribution of $Z$ given $Y$, respectively. We sample from
\begin{equation}\label{eq: posterior_predictive_rubin}
    p(Y_u \given y, Z) \propto \int p(Y_u \given y, Z, \theta, \phi)\times p(\theta, \phi \given y, Z)\mathop{d\theta}\mathop{d\phi}\;, 
\end{equation}
where $Y = \{Y_u, y\}$ with $y$ denoting observed samples and $Y_u$ unknown values in population units excluded in the sampling design. In order to sample from (\ref{eq: posterior_predictive_rubin}), we note that $p(Y_u, \theta, \phi \given y, Z) = p(Y_u \given y, Z, \theta, \phi)\times p(\theta, \phi \given y, Z)$. Therefore, we first sample values of $\theta$ and $\phi$ from their joint posterior distribution $p(\theta, \phi \given y, Z)$ and, subsequently, we draw one value of $Y_u$ from $ p(Y_u \given y, Z, \theta, \phi)$ using each sampled value of $\{\theta,\phi\}$. 

\subsection{Ignorable sampling designs}\label{sec: ignorable_sampling_designs}

A crucial observation is that the conditional distribution $p(Z\given Y, \phi)$ \emph{models} the sampling design and is accounted for in the finite population inference. For example, in the models considered in Sections~\ref{sec: srswor_bhm}--\ref{sec: bayesian_multistage} $p(Z \given y, \phi) = p(Z)$ does not depend on the sampled values and there are no unknown design parameters $\phi$. Only the sizes of the population and sample are used, which means that information about the design does not transmit to inference on the unknown population values and, hence, these designs are referred to as \emph{ignorable}. As seen in the aforementioned sections, purely model-based approaches can reproduce the Horvitz-Thompson estimator without requiring any information on the sampling design.      

Inference for finite population units from an ignorable design is, therefore, essentially a model-based prediction or Bayesian imputation problem. Linear regression models with dependent population units modify (\ref{eq: conjugate_bayesian_linear_regression_blocks}) and formulated as
  \begin{equation}\label{eq: bayesian_linear_regression_dependent}
  \begin{split}
      \begin{bmatrix} y \\ Y_u \end{bmatrix} &= \begin{bmatrix} X_s \\ X_u \end{bmatrix}\beta +  \begin{bmatrix} \epsilon_s \\ \epsilon_u \end{bmatrix}\;;\quad \begin{bmatrix} \epsilon_s \\ \epsilon_u \end{bmatrix} \sim N\left(\begin{bmatrix} 0 \\ 0 \end{bmatrix}, \begin{bmatrix} V_{s}(\theta) & V_{su}(\theta) \\ V_{us}(\theta) & V_{u}(\theta) \end{bmatrix}\right)\;;\\
 \beta &= A\mu + \eta\;;\quad \eta \sim N(0,V_{\beta})\;;\quad \theta \sim p(\theta)\;,
  \end{split}
 \end{equation}   
where $X_s$ and $X_u$ are fixed design matrices corresponding to sampled and non-sampled units, respectively, $\begin{bmatrix} V_{s}(\theta) & V_{su}(\theta) \\ V_{us}(\theta) & V_{u}(\theta) \end{bmatrix}$ is a positive definite covariance matrix partitioned accordingly, and $A$ and $\mu$ are assumed fixed. Inference proceeds from the following steps: (i) Draw samples from the posterior distribution $p(\beta, \theta \given y)$; (ii) Draw samples from $p(Y_u\given y)$ so that $\mathbb{E}_{\beta,\theta\given y} \left[N\left(Y_u\given X_u\beta + V_{us}V_{s}^{-1}(y - X_s\beta), V_u - V_{us}V_s^{-1}V_{su}\right)\right]$. The posterior samples $Y_u$ and the observations $y$ are then substituted into desired functions of the population units to obtain the posterior samples of the desired functions. The posterior estimator obtained from (\ref{eq: bayesian_linear_regression_dependent}) is consistent in the sense that $\mathbb{E}[Y_u \given y, \beta, \theta] = X_u\beta + V_{us}V_{s}^{-1}(y - X_s\beta)$ equals $y$ if the entire finite population is sampled whereupon $X_u = X_s$ and $V_{us} = V_s$ and the expression becomes free of model parameters $\{\beta,\theta\}$. Furthermore, the conditional variance $\mathbb{V}(Y_u \given y, \beta, \theta)$ equals $0$ when the finite population is sampled implying that the posterior estimator equals the finite population quantity with no uncertainty. In spatial contexts, which we discuss below, this reveals an interpolation property for finite populations using (\ref{eq: bayesian_linear_regression_dependent}).

\subsection{Spatial structured populations}\label{sec: spatial_pop}
A contemporary problem in survey sampling concerns estimation in spatially oriented finite populations, where each unit emerges as a (partial) realisation of a spatial process. In this domain, there is a substantial literature on small area estimation for regionally aggregated data \citep[see, e.g.,][]{Rao:vz, Clayton1987:eb, datta1991bayesian, Ghosh1994:sa, arora1997sinica, Ghosh1998:kn}, where interest lies in modelling dependencies across regions such as counties or districts within states, states within a country, or census-tracts or postal codes within a state or city. The population units correspond to regions are assumed to be correlated through a Markov Random Field positing that each region is conditionally dependent only on its neighbours, given all other regions. Conditional Auto-Regression (CAR) models are popular in small area estimation to model such dependence; see the aforementioned literature.  

Modelling considerations differ from CAR models when we consider quantities that, at least conceptually, exist in continuum over the entire domain, although they are recorded at a finite number of point-referenced locations. Now, dependence among the population units is induced from a spatial process, which assigns a probability law to an uncountable subset within a $d$-dimensional Euclidean domain. In general, spatial process modelling (\citealt{Banerjee:2014wm}; \citealt{Cressie:2011uu}; and \citealt{Ripley:2004ut}) follows the generic paradigm
\begin{equation}\label{eq: generic_paradigm}
[\mbox{data}\given \mbox{process}] \times [\mbox{process}\given \mbox{parameters}] \times [\mbox{parameters}]\;,
\end{equation}
which accommodates complex dependencies and multiple sources of variation. The literature on finite population sampling in spatial process settings appears to be considerably sparser than small area estimation. Here, \cite{VerHoef:2002xy} discuss connections between geostatistical models and classical design-based sampling and develop methods for executing model-based block ``kriging''. \cite{Cicchitelli:2012me} present a spline regression model-assisted, design-based estimator of the mean for use on a random sample from both finite and infinite spatial populations, while \cite{Bruno2013:fb} use linear spatial interpolation to create a design-based predictor of values at unobserved locations. 

For ignorable designs, the framework in (\ref{eq: bayesian_linear_regression_dependent}) is often sufficient. Specifically in spatial contexts, the elements of the covariance matrix in (\ref{eq: bayesian_linear_regression_dependent}) are the values of a positive definite covariance function. Several choices for the correlation function are available \citep[see, e.g.][]{Banerjee:2014wm}. The isotropic Mat\'ern function, which models spatial covariance as a function of distance between spatial locations, while also introducing a parameter to control smoothness of process realisations, is defined as $C(d_{ab}) = \sigma^2 + \tau^2$ if $d_{ab}=0$ and $C(d_{ab}) =  \tau^2 \frac{2^{1-\eta}}{\Gamma(\eta)}\left(\sqrt{2\eta}d_{ab}\phi\right)^\eta K_\eta \left(\sqrt{2\eta}d_{ab}\phi\right)$ if $d_{ab} > 0$, where $K_{\eta}(\cdot)$ is the modified Bessel function and $d_{ab}$ is the distance between two locations $\ell_a$ and $\ell_b$. The process parameters $\theta = \{\sigma^2, \tau^2, \phi, \eta\}$ capture variation due to measurement error or micro-resolution variation, variation attributable to spatial structure, the rate of decline in spatial association, and smoothness of the process, respectively. The exponential covariance function $C(d_{ab}) = \tau^2\exp(-\phi d_{ab})$ is a special case of Mat\'ern when $\eta = 1/2$.

\subsection{Non-ignorable spatial designs}\label{sec: nonignorable_spatial_designs}

A more general framework for spatial survey sampling that accommodates non-ignorable sampling designs is built from the approach surrounding (\ref{eq: posterior_predictive_rubin}). We introduce some further notations. Formally, define a spatial domain ${\cal L} \subseteq \mathbb{R}^2$ and let ${\cal L}_{FP} = \{{\ell}_1, \dots, {\ell}_N\}$ denote a finite collection of $N$ spatial locations. This set represents the units in our finite population and is made up of locations that are sampled, denoted by ${\cal L}_s$, and those that are not, denoted by ${\cal L}_u$; therefore, ${\cal L}_s \cup {\cal L}_u = {\cal L}_{FP}$. In addition, certain spatial applications will require a ``structural zero'' subset, denoted by ${\cal L}_{0} \subset {\cal L} \setminus {\cal L}_{FP}$, which represents all locations where sampling is precluded due to geographic considerations. For example, points situated over lakes and rivers may be precluded from the finite population of interest when one is interested in household economic surveys. Likewise, heavily urbanised metropolitan areas are precluded from studies concerning biomass interpolation from a finite population of trees in a forest.   

It is of interest to learn about the process in (\ref{eq: generic_paradigm}) from the sampled observations. A spatial process endows a probability law over the uncountable collection of random variables $\{(Y(\ell), Z(\ell)) : \ell \in {\cal L}\}$ and, hence, induces one for the finite population $\{(Y(\ell), Z(\ell)) : \ell \in {\cal L}_{FP}\}$. We consider three underlying stochastic processes generating the data: (i) a possibly vector-valued latent spatial process, $\{w(\ell): \ell \in {\cal L}\}$, which models the variable(s) of interest at $\ell$; (ii) a sampling indicator process, $\{Z(\ell): \ell \in {\cal L}\}$, which indicates whether a location is sampled or not; and possibly (iii) a structural zero indicator process (partially observed), $\{u(\ell): \ell \in {\cal L}\}$, which indicates whether $\ell \in {\cal L}_{FP}$ or not. A joint hierarchical model, conceptually, is constructed as
\begin{equation}\label{eq: bayesian_hierarchical_noningnorable_generic}
    \begin{split}
     [Y(\cdot)\given w(\cdot), u(\cdot)] \times [Z(\cdot)\given Y(\cdot), u(\cdot)] \times [w(\cdot) \given u(\cdot)]\times [u(\cdot)]\;.  
    \end{split}
\end{equation}
We turn to specifying probability models for each of these components in the spirit of \cite{finley2011jasa}, who predicted continuous forest variables at new locations accounting for uncertainty in whether the new locations were forested or not. 

\subsubsection{Modelling $[Y(\cdot)\given w(\cdot), u(\cdot)]$}
Let $u(\ell) = 1$ if $\ell \in {\cal L}_{FP}$ and $u(\ell) = 0$ if $\ell \notin {\cal L}_{FP}$. The first component in (\ref{eq: bayesian_hierarchical_noningnorable_generic}) is modelled as a customary spatial (or spatial-temporal) regression model for point-referenced data if $u(\ell) = 1$ and is distributed as the Dirac measure, $\delta_0$, with point mass at $0$ if $u(\ell) = 0$. Thus,   
\begin{align*}%\label{eq: generic_paradigm}
& Y(\ell) = u(\ell)\{x(\ell)^{\T}\beta + \tilde{x}(\ell)^{\T}w(\ell) + \epsilon(\ell)\} + (1-u(\ell))\delta_0\;;\quad \epsilon(\ell) \stackrel{iid}{\sim} N(0,\sigma^2)
 \end{align*}
where $x(\ell)$ is a vector of explanatory (perhaps including design) variables at $\ell$ with regression coefficients $\beta$, $\tilde{x}(\ell)$ is a subset of $x(\ell)$ whose regression coefficients, $w(\ell)$, vary over space, that is, the impact of the predictors on the population units vary over space and $w(\ell)$ is either a multivariate Gaussian process with a valid cross-covariance matrix or a vector of independent univariate processes \citep[see, e.g.,][]{Banerjee:2014wm} or using matrix-variate normal distributions \citep{zhang:2022sf}. The white noise process $\epsilon(\ell)$ captures micro-scale variation existing at fine scales (smaller than the minimum distances among locations in the finite population) that is attributed to measurement errors in surveys.      

\subsubsection{Modelling $[Z(\cdot)\given Y(\cdot), \omega(\cdot), u(\cdot)]$}
While the sampling indicator process $Z(\cdot)$ ``disappears'' from the likelihood or the posterior for ignorable designs, we can model non-ignorable designs using such a process. For example, we consider the model  
\begin{align*}%\label{eq: generic_paradigm}
& Z(\ell) \sim u(\ell)\mbox{Ber}(\pi(\ell)) + (1-u(\ell))\delta_{0}\;;\quad \mbox{logit}(\pi(\ell)) = \beta_{0Z} + \beta_{1Z} Y(\ell) + x_{Z}^{\T}\beta_{[2:p]Z} \;,
%& \mbox{logit}(\pi(\ell)) = z(\ell)\{\theta_{0} + \theta_{1} y(\ell)\} + (1-z(\ell))\times (-\infty) \;.
\end{align*}
where, for an arbitrary location $\ell$, the indicator process is distributed as a Bernoulli distribution with probability $\pi(\ell)$ if $u(\ell) = 1$, that is, if $\ell \in {\cal L}_{FP}$, and is set equal to zero if $u(\ell) = 0$, that is, if $\ell \notin {\cal L}_{FP}$. We introduce external predictors, $x_Z$, along with the finite population units $Y(\ell)$ to inform the sampling indicator process. In fact, one can conceive of introducing another spatial process
\begin{align*}%\label{eq: generic_paradigm}
& \mbox{logit}(\pi(\ell)) = \beta_{0Z} + \beta_{1Z} Y(\ell) + x_{Z}^{\T}\beta_{Z} + {\omega(\ell)} \;;\quad  \omega(\ell) \sim GP(0,C_{\theta_{\omega}}(\cdot))\;,
\end{align*}
where $\omega(\ell)$ is a zero centred Gaussian process with a specified covariance kernel $C_{\theta_{\omega}}(\cdot)$.  

\subsubsection{Modelling $[w(\cdot) \given u(\cdot)] = [w(\cdot)]$ and $[u(\cdot)]$}
The spatial process $w(\ell)$ is assumed to be independent of the population indicator $u(\ell)$ and is modelled as a multivariate Gaussian process. Let $w(\ell) = (w_1(\ell),\ldots, w_q(\ell))^{\T}$ be a $q\times 1$ vector, where $w_j(\ell)$ is a real-valued stochastic process over ${\cal L}$. The process is customarily specified by its mean $\mathbb{E}[w_j(\ell)] = \mu_j(\ell)$, which we assume to be zero, and the collection of stationary covariance functions $C_{ij}(h) = \mbox{Cov}(w_i(\ell), w_j(\ell + h))$, for $i,j=1,2,\ldots,q$. The $q\times q$ matrix $C(h) = (C_{ij}(h))$ is the \emph{cross-covariance} matrix and must satisfy (i) $C(h)^{\T} = C(-h)$ (note that $C(h)$ itself need not be symmetric), and (ii) $\sum_{i,j=1}^n a_i^{\T}C_{ij}(h_{ij})a_j \geq 0$ for any set of vectors $a_1,\ldots,a_q$ in $\mathbb{R}^q$, where $h_{ij} = \ell_i - \ell_j$, to ensure a valid multivariate spatial process for $w(\ell)$. A variety of methods are available for constructing multivariate spatial processes \citep{Banerjee:2014wm, gelban10, genton2015cross, zhang:2022sf}.

The quantity $u(\ell)$ is a population indicator process modelled by
\begin{align*}%\label{eq: generic_paradigm}
u(\ell) &\sim \mbox{Ber}(\pi(\ell))\;;\quad \mbox{logit}(\pi(\ell)) = x_{u}(\ell)^{\top}\beta_u + v(\ell)\;;\quad
v(\ell)  \sim GP(0, C_{\theta_v}(\cdot))\;,
\end{align*}
where $v(\ell)$ is a GP with covariance function $C_{\theta_v}(\cdot)$. Here, the model for $\pi(\ell) = P(u(\ell) = 1)$ is a logistic regression accommodating explanatory variables that inform about a location being a part of the finite population or not.  

\subsection{Graphical models for finite populations}\label{sec: graphical_superpopulations}
\begin{figure}[t]
\begin{center}
 \includegraphics[scale=0.35]{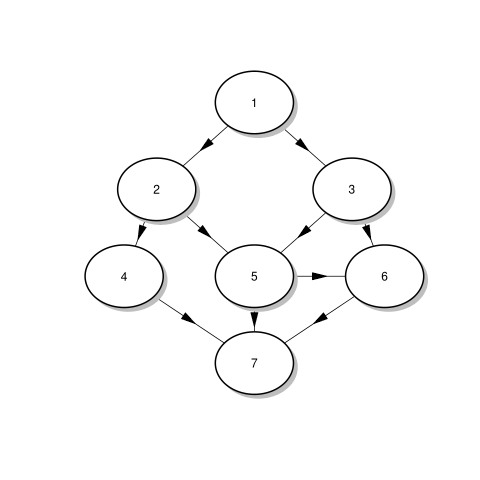}
 \includegraphics[scale=0.35]{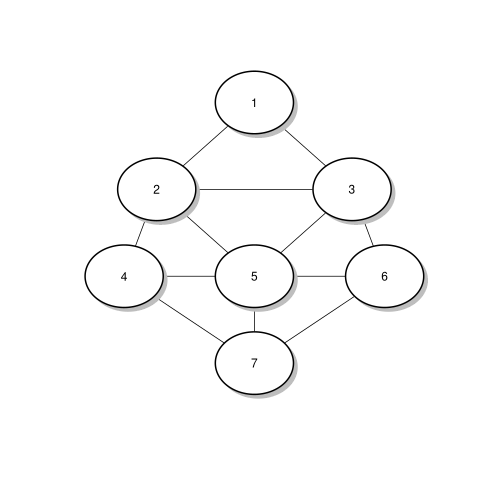}
\end{center}
\vspace{-0.15in}
\caption{Left: A directed graph (Bayesian network) with arrows pointing from parents to children. Right: An undirected conditional independence graph where the absence of edges between two nodes represent conditional independence given all other nodes.}\label{fig: graphs}
\end{figure}

The ability to endow probability laws over the finite population units makes the Bayesian framework substantially richer than design-based or classical model-based frameworks. For example, we introduce complex multivariate dependencies  among the finite population units using graphical models. Such models are based upon joint probability distributions over directed acyclic graphs (DAGs or Bayesian networks) and undirected graphs (left and right panels in Figure~\ref{fig: graphs}, respectively), where each node represents the value(s) of our variable(s) of interest from a population unit. An edge between two nodes in an undirected graph indicates that the random variables from the two nodes are conditionally dependent given all other variables, while the absence of an edge implies conditional independence given the remaining variables. A DAG, due to its acyclic property, always has at least one source node (or a node that does not have an edge incident on it, hence has no parents) and one can always construct a sequence of ordered nodes (topological order) from its structure. Using this, the DAG defines a valid joint probability distribution as a product of the probability densities of the variable(s) from a node conditional on its parents.

Graphical models are especially significant in contemporary multivariate data analysis \citep{cox1996graph, cox1993graph}. Let $Y_i = (Y_{i1},\ldots, Y_{iN})^{\T}$ be an $N\times 1$ random vector with elements $Y_{ij}$ denoting the values of a variable of interest indexed by $i=1,2,\ldots,q$ over population units $j=1,2,\ldots,N$. Let ${\cal G} = \{{\cal V}, {\cal E}\}$ be a posited undirected graph, where ${\cal V}$ is the set of $q$ vertices representing the variables and ${\cal E}$ is the set of all edges between pairs of variables that are assumed to be conditionally dependent with each other given all other variables. We specify the following sequence of full conditional distributions for modelling $Y_i$'s given all other $Y_j$'s for $j \in \{1,2,\ldots,q\}\setminus \{i\}$:
\begin{equation}\label{eq: full_conditionals}
 Y_i \given Y_{(-i)} \sim N\left(\sum_{j=1}^N A_{ij}Y_j, \Gamma_i\right)\;,\quad i=1,2,\ldots,q\;,
\end{equation}
where $Y_{-(i)} = \{Y_1, Y_2,\ldots, Y_q\}\setminus \{Y_i\}$, i.e., the collection of $Y_j$'s for $j=1,2,\ldots,q$ but excluding $Y_i$, $A_{ij}$'s are fixed $N\times N$ matrices, $A_{ii} = O$ (the matrix of zeroes), and $\Gamma_i$'s are fixed positive definite matrices. Since each variable can exhibit its own dependence structure, the $\Gamma_i$ varies by variable. Brook's Lemma provides a straightforward method for deriving the joint density from (\ref{eq: full_conditionals}) using the identity:
\begin{multline}\label{eq: brooks_lemma}
 p(Y_1, Y_2,\ldots, Y_q) = \prod_{i=1}^q \frac{p(Y_i \given \tilde{Y}_1,\ldots,\tilde{Y}_{i-1}, Y_{i+1},\ldots, Y_q)}{p(
\tilde{Y}_i\given \tilde{Y}_1,\ldots,\tilde{Y}_{i-1}, Y_{i+1},\ldots, Y_q))}
\times p(\tilde{Y}_1, \tilde{Y}_2,\ldots, \tilde{Y}_q)\;,
\end{multline}
where $\tilde{Y} = (\tilde{Y}_{1}^{\T},\tilde{Y}_2^{\T},\ldots,\tilde{Y}_q^{\T})^{\T}$ is any fixed point in the support of $p(Y)$ and we assume that the joint density $p(\cdot) > 0$ over its entire support. The proof is a straightforward verification proceeding from the last element of the right hand side (i.e., the joint density on the right hand side). Thus, we begin with the observation that
\begin{align*}
\frac{p(Y_q \given \tilde{Y}_1,\ldots,\tilde{Y}_{q-1})}{p(\tilde{Y}_q\given \tilde{Y}_1,\ldots,\tilde{Y}_{q-1})} \times p(\tilde{Y}_1, \tilde{Y}_2,\ldots, \tilde{Y}_q)
 &= p(\tilde{Y}_1, \tilde{Y}_2,\ldots, \tilde{Y}_{q-1}, {Y}_q)\;. 
\end{align*}
\begin{comment}
{\footnotesize
\begin{align*}
   \frac{p(Y_q \given \tilde{Y}_1,\ldots,\tilde{Y}_{q-1})}{p(
\tilde{Y}_q\given \tilde{Y}_1,\ldots,\tilde{Y}_{q-1})} \times p(\tilde{Y}_1, \tilde{Y}_2,\ldots, \tilde{Y}_q)
 &= \frac{p(Y_q \given \tilde{Y}_1,\ldots,\tilde{Y}_{q-1})}{\cancel{p(
\tilde{Y}_q\given \tilde{Y}_1,\ldots,\tilde{Y}_{q-1})}} \times p(\tilde{Y}_1,\ldots, \tilde{Y}_{q-1}) \times \cancel{p(\tilde{Y}_q \given \tilde{Y}_1,\ldots, \tilde{Y}_{q-1})} \\
&= p(\tilde{Y}_1, \tilde{Y}_2,\ldots, \tilde{Y}_{q-1}, {Y}_q)\;.
\end{align*}
}
\end{comment}
\begin{comment}
This reduces the right hand side of (\ref{eq: brooks_lemma}) to
\[
\prod_{i=1}^{q-1} \frac{\pi(z_i \given \tilde{z}_1,\ldots,\tilde{z}_{i-1}, z_{i+1},\ldots, z_{q-1})}{\pi(
\tilde{z}_i\given \tilde{z}_1,\ldots,\tilde{z}_{i-1}, z_{i+1},\ldots, z_{q-1}))} \\
\times \pi(\tilde{z}_1, \tilde{z}_2,\ldots, \tilde{z}_{q-1}, z_q)\;.
\]
\end{comment}
Proceeding as above will continue to replace $\tilde{Y}_{i}$ with $Y_i$ in the joint density on the right hand side of (\ref{eq: brooks_lemma}) for each $i=q-1, q-2,\ldots, 1$ and we eventually arrive at $p(Y_1, Y_2,\ldots, Y_q)$. Applying (\ref{eq: brooks_lemma}) to the full conditional distributions in (\ref{eq: full_conditionals}) with $\tilde{Y}_i = 0$ for each $i=1,\ldots,q$ yields the joint density
\begin{equation}\label{eq: gmrf}
\begin{split}
  p(Y) &\propto \exp\left\{-\frac{1}{2}\left(\sum_{i=1}^q Y_i^{\T}\Gamma_i^{-1}Y_i - \sum_{i=1}^q\sum_{j\neq i}^q Y_i^{\T}\Gamma_i^{-1}A_{ij}Y_j\right)\right\} \propto \exp\left(-\frac{1}{2}Y^{\T}QY\right)\;,
 \end{split}
\end{equation}
where $Y = (Y_1^{\T},\ldots, Y_q^{\T})^{\T}$ is $Nq \times 1$ and $Q = M^{-1}(I-A)$ is $Nq \times Nq$, $M = \oplus \Gamma_i$ is block-diagonal with $(i,i)$-th block $\Gamma_i$ for $i=1,\ldots,q$ and $A = (A_{ij})$ is the $Nq \times Nq$ block matrix with $A_{ij}$ as the $(i,j)$th block. For (\ref{eq: gmrf}) to be a valid density, $Q$ needs to be symmetric and positive definite so that $Y \sim N(0, Q^{-1})$. In addition, $Q$ must conform to the conditional independence relationships in ${\cal G}$. To be precise, if two distinct nodes $i$ and $j$ in the graph do not have an edge, then $Y_i \perp Y_j \given Y_{-(i,j)}$ or, equivalently, the $N\times N$ block of the precision $Q_{ij} = -\Gamma_i^{-1}A_{ij} = O$. Since $A_{ii}=O$ in (\ref{eq: full_conditionals}), the $(i,i)$-th block of $Q$ is $\Gamma_{i}^{-1}$. We devise one such construction below. 

Given the inter-variable graph ${\cal G} = \{{\cal V}, {\cal E}\}$ let $\Lambda = (\lambda_{ij}) = D - \rho W$ be the $q\times q$ graph Laplacian, where $W=(w_{ij})$ is the adjacency matrix with nonzero $w_{ij}$ only if there is an edge between $i$ and $j$, $D=(d_{ii})$ is a $q\times q$ diagonal matrix with the sum of each row of $W$ along the diagonal, i.e., $d_{ii} = \sum_{j=1}^q w_{ij}$ , and $\rho$ is a scalar parameter that ensures positive-definiteness of $\Lambda$ as long as $\rho \in (1/\zeta_{min}, \zeta_{max})$, where $\zeta_{min}$ and $\zeta_{max}$ are the minimum and maximum eigenvalues of $D^{-1/2}WD^{-1/2}$, respectively. Let $R_i$ be any nonsingular $N\times N$ factor (e.g., a triangular Cholesky) of the positive definite precision matrix of $Y_i$, i.e., $(\mbox{var}(Y_i))^{-1} = C_{ii}^{-1} = R_i^{\T}R_i$ for each $i=1,2,\ldots,q$. We set $\Gamma_i^{-1} = \lambda_{ii}R_i^{\T}R_i$ and $A_{ij} = -(\lambda_{ij}/\lambda_{ii})R_i^{-1}R_j$ in (\ref{eq: full_conditionals}). Then $Q_{ij} = \lambda_{ij}R_{i}^{\T}R_j$ and $Q = \tilde{R}^{\T}(\Lambda \otimes I)\tilde{R}$, where $\tilde{R} = \oplus_{i=1}^q R_i$. Since each $R_i$ is nonsingular and $\Lambda$ is positive definite, it follows that $Q$ is positive definite and, hence, $\det(Q) = \prod_{i=1}^q (\det(R_i))^2(\det(\Lambda))^n$. The above construction yields proper densities in (\ref{eq: gmrf}) that will conform to conditional independence among the variables represented by the nodes of a posited undirected graph and can also be adapted for analysing high-dimensional multivariate data, where at least one of or both $N$ and $q$ are large. The special case where each $R_i=R$ for $i=1,2,\ldots,q$ yields the separable model $Q = \Lambda \otimes (R^{\T}R)$.   

If, for example, the population units are spatially structured, as is often the case in small area estimation, then spatial dependencies are introduced in the $R_i$'s. The CAR model, which is a popular choice in small area estimation, uses the Laplacian of a spatial graph, say ${\cal G}_s = \{{\cal V}_s, {\cal E}_s\}$, for modelling spatial concentrations. The vertices in ${\cal V}_s$ are the population units and the edges in ${\cal E}_s$ between two units indicate they are spatial neighbours. The suffix $s$ is used to distinguish this graph from the inter-variable graph in the previous paragraph. Each $R_i$ satisfies $R_i^{\T}R_i = D_s - \rho_{s,i} W_s$, where $D_s$ is $N\times N$ diagonal with the number of neighbours for each region as its diagonal elements and $W_s$ is the $N\times N$ spatial adjacency matrix. The parameter $\rho_{s,i}$ depends on the variable index $i$ because each variable is allowed to have its own spatial correlation parameter. Since $D_s^{-1/2}WD_s^{-1/2}$ is symmetric, there is an orthonormal basis of its eigenvectors $\{u_1, u_2,\ldots, u_N\}$ in $\mathbb{R}^N$ consisting of its eigenvectors, and we can write $I_N - \rho_{s,i}D_s^{-1/2}WD_s^{-1/2} = \sum_{j=1}^N (1-\zeta_{s,j}\rho_{s,i})u_ju_j^{\T}$, where $\zeta_{s,j}$ is the eigenvalue corresponding to $u_j$. Since ${\cal G}_s$ is fixed from the map of population units, the eigenvalues and eigenvectors are computed only once and fixed. Each spatial concentration parameter $\rho_{s,i}$ is restricted to be in an interval so that $1-\zeta_{s,j}\rho_{s,i} > 0$ for each $j=1,2,\ldots,N$. If $R_i = \sum_{j=1}^N \sqrt{(1-\zeta_{s,j}\rho_{s,i})}u_ju_j^{\T}/\sqrt{d_{s,j}}$, where $d_{s,j}$ is the $j$-th diagonal element of $D_s$, then $R_i^2$ is the symmetric square root of the spatial concentration matrix and depends only on one unknown parameter, $\rho_{s,i}$, which is estimated from its posterior.            

\section{Model assessments and comparisons}\label{sec: model_assessment}
The preceding discussions demonstrates the richness of Bayesian hierarchical modelling for estimating finite population quantities. This richness comes at an expense of justifying the choice of the model for the population, which requires using Bayesian model selection metrics. For example, consider the two models (\ref{eq: bayesian_ht_little}) and (\ref{eq: bayesian_ht_sinha_ghosh}). Both these models emulate (one approximate and the other exact) the Horvitz-Thompson estimator, but differ in their variance structure. A design-based perspective favouring the Horvitz-Thompson estimator focuses primarily on its unbiased nature and on estimating its variance. A model-based perspective, including a fully Bayesian perspective, entails deciding between these models based upon a realised sample.       

To elucidate further, assume that we wish to compare two models within the framework of (\ref{eq: bayesian_linear_regression_dependent}) but with possibly different specifications for the design matrix and the dependence structure. A key quantity in Bayesian model selection is the point-wise predictive distribution \citep[see, e.g.,][]{Gelman:2014tc} defined as
\begin{equation}\label{eq: post_pred}
    p(Y_{rep,i}\given y) = \mathbb{E}_{\beta,\theta\given y}\left[p(Y_{rep,i}\given \beta, \theta)\right] = \int p(Y_{rep,i} \given \beta, \theta) \times p(\theta, \beta \given y) \mathop{d\beta} \mathop{d\theta} 
\end{equation}
for each unit $i$ in the population. We sample one value of $Y_{rep,i} \sim p\left(Y_{rep,i} \given \beta, \theta \right)$ for each posterior sample of $\beta$ and $\theta$. An ideal measure of a model's performance is its out-of-sample predictive performance for new data produced from the true data-generating process. This, unfortunately, is impracticable (except in synthetic experiments that simulate the population) because we do not know the true data generating process. Therefore, the predictive distribution (\ref{eq: post_pred}) encodes the distribution of alternative data that would be replicated from the fitted model.       

\cite{Gelfand:1998wj} uses the predictive distribution to define
\begin{equation}\label{eq: gelfand_ghosh}
    D = \sum_{i\in {\cal S}} (y_i -\mathbb{E}[Y_{rep,i} \given {y}])^2 + \sum_{i \in {\cal S}} \mathbb{V}(Y_{rep,i}\given {y})
\end{equation}
as a model choice metric, while \cite{Gneiting2007:pq} prefer 
\begin{equation}\label{eq: grs}
 GRS = -\sum_{i \in {\cal S}} \frac{(y_i -\mathbb{E}[Y_{rep,i} \given {y}])^2}{\mathbb{V}(Y_{rep,i}\given {y})} - \sum_{i \in {\cal S}} \log \mathbb{V}(Y_{rep,i}\given {y})\;,   
\end{equation}
as a proper scoring rule, where ${\cal S}$ is the set of sampled units. Lower values of $D$ and higher values of $GRS$ indicate better model fit. A third, more recently developed, metric is the Watanabe-Akaike Information Criterion \citep[see, e.g.,][]{watanabe2013:a, Vehtari:2017wy}, which computes the logarithm of (\ref{eq: post_pred}) evaluated at each data point and adjusts with a penalty term as
\begin{equation}\label{eq: waic}
    WAIC = -2\sum_{i\in {\cal S}} \log \mathbb{E}_{\beta,\theta \given y}[p(y_{i}\given \beta, \theta)] + 2\sum_{i\in {\cal S}}\mathbb{V}_{\beta,\theta\given y}\left(\log p(y_{i}\given \beta,\theta)\right)\;.
\end{equation}
The $WAIC$ computed as above is in the same scale as the deviance with smaller values indicating superior performance.

Each of the metrics (\ref{eq: gelfand_ghosh})---(\ref{eq: waic}) are computed without requiring any information on a true data generating model. The quantities (\ref{eq: gelfand_ghosh})~and~(\ref{eq: grs}) are computed using samples of $Y_{rep,i}$ from (\ref{eq: post_pred}) to calculate $\mathbb{E}[Y_{rep,i}\given y]$ and $\mathbb{V}(Y_{rep,i}\given y)$. The remaining information used comes from the realised sample of values from the finite population. We compute the quantities in (\ref{eq: waic}) using $\sum_{l=1}^L p(y_i \given \beta^{(l)}, \theta^{(l)}) / L$ and $\mbox{var}_{l=1,\ldots,L}(\log p(y_i \given \beta^{(l)}, \theta^{(l)}))$ to estimate $\mathbb{E}_{\beta,\theta \given y}[p(y_{i}\given \beta, \theta)]$ and $\mathbb{V}_{\beta,\theta\given y}\left(\log p(y_{i}\given \beta,\theta)\right)$, respectively, where $\mbox{var}_{l=1,\ldots,L}(a^{(l)}) = \sum_{l=1}^L \left(a^{(l)} - \left(\sum_{l=1}^L a^{(l)}/L\right)\right)^2/(L-1)$ and ${\beta^{(l)}, \theta^{(l)}}$ are posterior samples $l=1,\ldots,L$. Vehtari et al. \citep{Vehtari:2017wy} provide further details on $WAIC$, including methods to compute standard errors of $WAIC$ \citep[used by][in finite population inference]{chan-golston:2020ab} and its connections to predictive leave-one-out (LOO) comparisons.     
 
\section{Examples}\label{sec: examples}
I present some salient features of an analysis from two examples. The first is a study of ground-water nitrate levels in the population of California Central Valley wells situated in the Tulare Lake Basin (TLB), where health officials seek estimates of mean nitrate levels. The data we consider here has been extracted from the California Ambient Spatio-Temporal Information on Nitrate in Groundwater (CASTING) Database, which is described in \cite{Harter:2017vb} and \cite{Boyler:2012wv}. The spatial locations of these wells are available. Health implications of high nitrate levels in wells are dire. At high levels, infants and pregnant women are more susceptible to nitrate poisoning, which makes it more difficult for oxygen to be distributed to body and can be fatal to infants less than six months old. 

Our finite population consists of 6,117 unique wells among 63 zip codes. We follow a two-stage cluster sampling design described in \cite{chan-golston:2020ab}, where 21 zip codes were randomly chosen in the first stage and 50\%–90\% of the wells in each zip code were randomly sampled in the second stage. The nitrate levels in the sample wells were measured in milligrams per litre ( mg/L). The following models were used to analyse the sampled data: (i) the two-stage hierarchical cluster model described in Section~\ref{sec: bayesian_multistage}; (ii) a spatial random field model over the entire region that ignores the structure of the cluster sampling design (so a single set of spatial process parameters for the entire region); (iii) a two-stage model with an additive global spatial process; and (iv) a regional spatial process model where each cluster has its own mean parameter and spatial process. All of these models can, in fact, be subsumed into (\ref{eq: bayesian_linear_regression_dependent}) with appropriately constructed design and covariance matrices. 

Table~\ref{tab:da} presents the posterior mean and 95\% credible intervals for the finite population means obtained from each model. There are some discrepancies in these estimates among the different models. Hence, it will be appropriate to  The WAIC scores for each of the models are also presented. Some key features are worth pointing out. Notably, models that either ignore the two-stage structure of the sampled data (models ii and iv) or that ignore the spatial dependence (model i) appear to perform considerably worse (in terms of WAIC) than the model that accommodates both the sampling structure and the spatial dependence (model iii). Our recommendation is to report the analysis from model (iii). Also, while the credible intervals around the finite population mean from different models largely overlap, the point estimates from (i) and (iv) are lower than (ii) or (iii). This is, perhaps, attributed to the former two models not having a single spatial random field over the entire domain. Finally, the spatial clustering introduced in the latter two models result in shorter credible intervals.     

\begin{table}[t]
\caption{Estimates from a spatial meta-analysis California Nitrate Data Analysis}
\centering
\begin{tabular}{lll}
  \hline
 Model & Finite Population Mean in mg/L (95\% CI) & WAIC \\ 
  \hline
  (i) Two-Stage & 26.6 (19.2, 35.1) & 4701.6 \\ 
  (ii) Spatial & 32.6 (27.2, 38.7)  & 4858.4 \\ 
  (iii) Two-Stage + Spatial & 31.2 (24.2, 37.3) & 2697.0 \\ 
  (iv) Regional Spatial & 26.2 (18.7, 34.4) & 4536.6 \\ 
   \hline
\end{tabular}
\bigskip
\label{tab:da}
\end{table}

Unlike the above analysis, which assumed an ignorable design, I now present some key aspects of a second analysis from a study to estimate the percentage of annual reported income spent on fruits and vegetables (PIFV) in two low-income communities in Los Angeles, California \citep[see][and references therein for a detailed description of the study]{chan-golston:2022ab}. This study follows the framework described in Section~\ref{sec: nonignorable_spatial_designs}. There are $N=635$ locations in ${\cal L}_{FP}$ and within each location we have a number of households comprising the population units. The total number of population units is $2015$. We randomly sampled $n=555$ locations and then sampled households within each of these locations to obtain a sample of 1294 population units. These units reported amounts spent on fruits and vegetables on weekly, bi-weekly, or monthly scale. These values were multiplied by 52, 26, and 12, respectively, to reflect the annual amount spent on fruits and vegetables in a household. Reported yearly incomes in the population units were also collected. 

We consider a simpler version of the model presented in Section~\ref{sec: nonignorable_spatial_designs} and its subsections with the process $u(\cdot) = 1$. We assume that the PIFV of the $j$-th household within location $\ell_i$ is modelled as $Y_j(\ell_i) = x_j(\ell_i)^{\T}\beta + w(\ell_i) + \epsilon(\ell_i)$, $x_j(\ell_i)$ is a vector of fixed explanatory variables, $w(\ell_i)$ is a spatial random effect and $\epsilon(\ell_i)$ is white noise representing measurement errors and micro-scale variation. This is the model for $[Y(\cdot)\given w(\cdot)]$ in Section~\ref{sec: nonignorable_spatial_designs} with $u(\ell) = 1$. Based upon this, we consider four models: (i) a simple linear regression model that sets $w(\ell) = 0$ for all locations (yields an ignorable response); (ii) a model with nonignorable response that models $[Z(\cdot)\given Y(\cdot), \omega(\cdot)]$ in addition to $[Y(\cdot)\given w(\cdot)]$, but with no spatial effect (hence, $w(\cdot) = 0$ and $\omega(\cdot) = 0$); (iii) spatial random effects in $[Y(\cdot)\given w(\cdot)]$, but no spatial effects in $[Z(\cdot) \given Y(\cdot)]$; and (iv) fully spatial model with nonignorable responses with spatial effects $w(\cdot)$ and $\omega(\cdot)$ as described in Section~\ref{sec: nonignorable_spatial_designs}.

Table~\ref{t:da1} presents the posterior estimates of the finite population PIFV (first row) from the four models described above. The performance of these models was evaluated using $D$ and $GRS$ defined in (\ref{eq: gelfand_ghosh}) and (\ref{eq: grs}), respectively. Based upon these metrics, model~(iii) is marginally preferred to model~(iv), which suggests a lack of spatial dependence in the sampling inclusion model. Both the spatial models, (iii)~and~(iv), considerably outperform the non-spatial models including (ii) which still accommodates the nonignorable response. Model~(i), which assumes ignorability, performs the worse. The credible interval from the ignorable model seems very tight, while those from the other models are wider because of the propagation of uncertainty from the additional parameters being estimated.  The PIFV from model~(iii) is recommended for reporting given the model's superior performance.     

\begin{table}[t]
\caption{Results of regression models predicting percentage of income spent on fruits and vegetables (log-scale). The finite population mean income is presented in units of \$ 10,000.}\label{t:da1}
\begin{tabular}{lllll}
 \hline
 & Model 1 & Model 2 & Model 3 & Model 4 \\ 
 \hline
%Finite Pop. Avg. Inc. & 40.85 (37.19, 45.86) & 29.36 (26.66, 32.58) & 25.58 (24.14, 27.71) & 28.72 (26.18, 31.83) \\
PIFV \% & 17\% (16\%, 19\%) & 26\% (22\%, 31\%) & 35\% (28\%, 43\%) & 26\% (22\%, 32\%) \\
$D$ & 3527.8 & 3692.2 & 3701.2 & 3482 \\ 
$GRS$ & -1841.2 & -1903.6 & -1863 & -1767.8 \\ 
  \hline
\end{tabular}
\end{table}

\section{Discussion}\label{sec: discussion}
The article presents Bayesian inference for finite populations using survey data with an emphasis on settings where the finite population units are posited to exhibit complex dependence structures. Spatial process models for the finite population units are devised in ignorable and nonignorable response settings to bring out the richness and flexibility of Bayesian inference adapted to finite population survey sampling. Bayesian inference for finite populations in stochastic process settings has received relatively limited attention and can form a part of future extensions to this paper. Of particular interest to spatial statisticians will be the connections between models for nonignorable spatial responses and the preferential sampling framework described by \cite{Diggle:2010ir}. Modelling massive spatially oriented survey data is gaining attention in environmental sciences, ecology and forestry \citep[see, e.g.,][for a case study in forestry that compares Bayesian spatial models with design-based estimates for biomass and demonstrates the inferential benefits of the former]{finley2024jabes}. New methods for analysing massive spatially referenced survey data can adapt, perhaps even simplify, spatial meta-kriging approaches \citep{Guhaniyogi:2018ab} to finite population inference for ignorable and non-ignorable settings. Finally, attention is needed in the development and dissemination of software products for implementing Bayesian models for spatially oriented survey data. 

\bmhead{Acknowledgments}
The author wishes to thank the editors and an anonymous referee for valuable feedback. The author is especially grateful to Professors Roderick J. Little and Trivellore Raghunathan from the University of Michigan, Ann Arbor, U.S.A., and Professor J.N.K. Rao from Carleton University, Ottawa, Canada, for insightful discussions on inference for finite populations. The work of the author has been supported, in part, by the National Science Foundation (NSF) from grants NSF/DMS 1916349 and NSF/DMS 2113778, by the National Institute of Environmental Health Sciences (NIEHS) from grants R01ES030210 and R01ES027027 and by the National Institute of General Medical Science from grant R01GM148761.
    
\section*{Disclosures and declarations}

%Some journals require declarations to be submitted in a standardised format. Please check the Instructions for Authors of the journal to which you are submitting to see if you need to complete this section. If yes, your manuscript must contain the following sections under the heading `Declarations':

\begin{itemize}
\item Funding: The work of the author has been supported, in part, by the National Science Foundation (NSF) from grants NSF/DMS 1916349 and NSF/DMS 2113778, by the National Institute of Environmental Health Sciences (NIEHS) from grants R01ES030210 and R01ES027027 and by the National Institute of General Medical Science from grant R01GM148761.

\item Conflict of interest: The author declares that there are no financial or non-financial conflicts of interest.

%\item Ethics approval 
%\item Consent for publication
%\item Availability of data and materials
%\item Authors' contributions
\end{itemize}

\bibliography{sn-bibliography}% common bib file

\end{document}